\documentclass[preprint]{vgtc}                          
\preprinttext{This manuscript was  presented at alt.VIS, a workshop co-located with IEEE VIS 2021 (held virtually)}

\ifpdf
  \pdfoutput=1\relax                   
  \usepackage{graphicx}                
  \DeclareGraphicsExtensions{.pdf,.png,.jpg,.jpeg} 
\else
  \usepackage{graphicx}                
  \DeclareGraphicsExtensions{.eps}     
\fi%

\graphicspath{{figures/}{pictures/}{images/}{./}} 

\usepackage{microtype}                 
\PassOptionsToPackage{warn}{textcomp}  
\usepackage{textcomp}                  
\usepackage{mathptmx}                  
\usepackage{times}                     
\usepackage{cite}                      
\usepackage{tabu}                      
\usepackage{booktabs}                  
\usepackage{xcolor}
\usepackage[colorlinks=true,
    linkcolor=blue]{hyperref}
    
\newcommand{\MYhref}[3][blue]{\href{#2}{\color{#1}{#3}}}%

\onlineid{4940}

\vgtccategory{Research}

\vgtcinsertpkg

\title{Xenakis: Experimenting with Data, Cities, and Sounds}

\author{Victor Schetinger \thanks{e-mail: victor.schetinger@tuwien.ac.at} \\
        \scriptsize TU Wien %
\and Ignacio Pérez-Messina\thanks{e-mail: ignbpm@gmail.com}\\ %
     \scriptsize TU Wien %
\and Renan Guarese\thanks{e-mail: renan.martins.guarese@student.rmit.edu.au}\\ %
     \scriptsize Royal Melbourne Institute of Technology %
\and Velitchko Filipov\thanks{e-mail: velitchko.filipov@tuwien.ac.at}\\ %
     \parbox{1.4in}{\scriptsize \centering  TU Wien}}

\teaser{
  \centering
  \includegraphics[width=\linewidth]{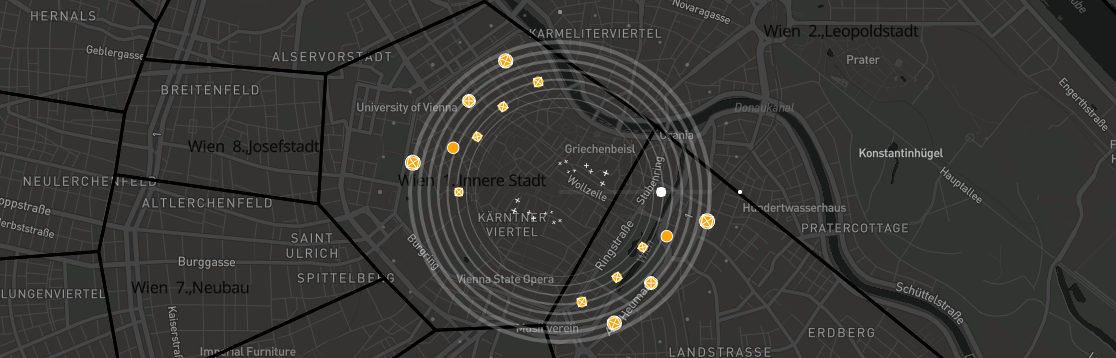}
  \caption{Vienna's streets summarized as a \MYhref{https://miro.com/app/board/o9J_l47ltks=/?moveToWidget=3074457361918048562&cot=14}{drum and bass} loop in our Xenakis music compass. GeoJSON street data is binned according to each street's orientation, capturing the textural features of the urban mesh and then turning them into sound. The compass is comprised of 32 beats (16 bins times 2), which could be unrolled into planar musical notation.}
  \label{fig:teaser}
}

\abstract{In this work, we report on the results and lessons learned from different disciplines while researching the loosely-defined problem of hearing a city. We present Xenakis, a tool for the musification of urban data, which is able to capture some features of a city's topology through the distribution of street orientations, and turn it into a (very) small piece of music, a loop, which can be used as building block for compositions. Besides providing complementary visual and auditory channels to interface with this data, we also allow the piping of \textit{midi} signals to other applications. This concept was developed by visualization researchers collaborating with musicians using design study methodologies in an open-ended way. Our results include musical tracks, and we take advantage of the scope of alt.VIS to communicate our research in a sincere, humorous, and engaging format.
} 

\CCScatlist{
  \CCScatTwelve{Human-centered computing}{Visu\-al\-iza\-tion}{Visu\-al\-iza\-tion Analytics}{Sonification};

}

\begin{document}


\firstsection{Introduction}

\maketitle


Most authors of this publication come strictly from a visualization background, and had no experience in developing sonification. Xenakis is the output of a continuous effort to test some ideas that seemed too good, but outside the reach of our comfort zone, both technically and theoretically. We wanted to hear what a city could sound like, not in a real sense, but based on some intrinsic features that, while visually noticeable, alluded our proper characterization. The expression of these features, we figured, was encoded in digital map representations because we could seem them when using, for instance, Google Maps. Therefore, we decided as starting point to treat this as a problem of visualizing GeoJSON data, but rendering it with sound.

``This is such as simple idea'', we thought, ``the web must be filled with similar things''. However, we did not know what keywords to search for, and our initial attempts at finding either a usable tool to listen to urban data, or a state-of-the-art publication on the subject were not fruitful. We decided to work on our own implementation of the concept, while simultaneously trying to map the state-of-the-art among different disciplines that brushed on the subject, among them data sonification, urban planning, and as the scope got broader soon we landed into musical territory. 

The jump from sound to music was a pretty obvious one: we don't want just to make sounds, we want them to sound good. And that is music. However, as visualization researchers it was anticipated that there could be a conflict between aesthetic criteria and clarity. The same way that a visualization uses visual encodings such as color, area, texture, and shape to convey information, sound also has its encoding variables: volume, pitch, position, timbre, and so on. If one is left to choose freely how to use them according to subjective criteria, it obfuscates the communication of information. But still, we want it to sound good. We want to have the cake and eat it, too.

 From a Visual Analytics point of view, the requirements of certain users and tasks can serve as excuse for breaking good practices. Therefore, we misappropriated design study methodologies to navigate this uncertain terrain. We approached musicians and proposed a role reversal: have them provide musical solutions to visualization problems rather than the usual, opposite. This of course creates a conflict of interest, short-circuiting the design study as we were both stakeholders and designers.  
 
When evaluating the quality of our output we realized it actually passed the rigor criteria of what is expected of a design study\cite{2019_infovis_criteria}: it was informed, plausible, abundant, resonant, and most of all reflexive. The last criteria, transparency, required a level of exposition that is normally not achievable in traditional academic media. Our contributions exist in the twilight zone between disciplines, and if they were scrutinized by the peer-review of a printable article, they would be crushed under \MYhref{https://miro.com/app/board/o9J_l47ltks=/?moveToWidget=3074457361941504825&cot=14}{disciplinary boundaries}. alt.VIS was part of our solution to have our cake and eat it, too. The weirdness clause of alt.VIS applies to the subject of research, the methodology, and also the presentation of this submission.

\subsection{Disclaimer: how to read this paper}

This paper presents an experiment in research, collaboration, and exposition. Part of the content is presented here in a double-column, abridged, and perhaps overly sincere scientific paper format. The other part is organized in a non-linear, topological fashion using \MYhref{https://miro.com/app/board/o9J_l47ltks=/?moveToWidget=3074457361915360658&cot=14}{a miro board}. Whenever blue text like this occurs, it is linking directly to an object inside the board. Once inside, the reader can follow any path they choose to navigate and explore the content. Different levels of zoom are used to nest information, and videos and images are weaved together to provide a hypermedia flow, which varies a bit according to each author's writing style. We tried to use only open-source or easily available link-able sources, applying the common sense of educational material. We apologize for any legal caveat that we might have missed that would make it wrong to share public knowledge.

\section{What Did We Actually Do}

Instead of having a methodology and results section, we decided to cut the chase and just tell what we actually did. Over the course of a few months, we had meetings among ourselves and with musicians, we sketched solutions, struggled to implement them, studied music production and composition, one of the authors had to start learning piano, and while all this was happening, we were investigating the literature of the adjacent fields. 

\subsection{What the Thing We Made Actually Does}

The current prototype has a map interface that starts centered on Vienna, but the user can go pretty much anywhere the Mapbox API has data on. In the center of the view, there is a \MYhref{https://miro.com/app/board/o9J_l47ltks=/?moveToWidget=3074457361917733112&cot=14}{"musical compass"} showing a histogram of street orientation distributions on the viewed area. In other words, it bins streets according to their angle, and since a north-south street and a south-north street are visually indistinguishable, the musical compass is symmetrical.

When the "Sonify Me!" button is pressed, this data is sent to a Chuck script that turns it into a loop pattern containing drums and a bass. Each bin corresponds to a beat, and the size of the beat determines the instruments and notes to be played, so that the higher the value of the bin, the more intense the sound is. A lot of effort was put in synchronizing the sound and visuals, both temporally spatially, so that the visualization and the sonification could be interchangeable for tasks that required a circular histogram encoding. Even more effort was put into respecting music theory fundamentals so that the experience of exploring musical configurations along the world could be engaging.

Rhythms that can be produced with our tool fall within a class of symmetrical rhythms with similar properties, based on the possible configurations of such type of circular histogram. Nevertheless, there is enough variety so that, even with these few elements, cities can sound very different for trained ears. A similar class of rhythms, called Euclidean rhythms\cite{toussaint2005euclidean}, has been shown to be present it different cultures' folk music, representing \MYhref{https://miro.com/app/board/o9J_l47ltks=/?moveToWidget=3074457361946173722&cot=14}{balanced distributions of pulses over time}

\subsubsection{So Just Street Orientations? What about buildings, parks...}

The current version of the prototype is focused on the street orientation problem to deliver a minimal viable product with a thoughtful music design. The data pipeline to musify anything present in the GeoJSON is already there, and we intend to develop this in future work. We explored different technical solutions for our \MYhref{https://miro.com/app/board/o9J_l47ltks=/?moveToWidget=3074457361926756757&cot=14}{architecture}, but our philosophy was: whatever is available, whatever works. 

\section{\MYhref{https://miro.com/app/board/o9J_l47ltks=/?moveToWidget=3074457361918760443&cot=144}{Related Work}}

Several authors have explored the use of sonification methods to display geolocated, cartographic or even urban-specific data to users \cite{KAKLANIS201359, Brittell:2013, Brittell:2018, Gune:2018, park_sihwa_2010_1177877, Parthenios:2016, Gaye2003, Adhitya:2011, Arango:2019, sarmento2020musical}. More often than not, these papers are fundamentally strict about their concept of sonification, tending to distance themselves from music, in the sense that sonification conveys meaning, rather than entertainment. According to Middleton et al. \cite{Middleton:2018}, their concern is that musical representations can create distractions by adding unnecessary elements for an analysis of the data, which is solely beneficial to form, in detriment of function.

On the other hand, Middleton et al. \cite{Middleton:2018} defends that musical elements can facilitate analytical objectives from improved data perception, data analysis, and data interpretation through user engagement and enhanced experiences. In this line of thought, a considerable amount of works has explored sonification of data via a musical perspective \cite{Allison2012, Arango:2019, BEARMAN2012157, Gaye2003, Middleton:2018, park_sihwa_2010_1177877, Parthenios:2016, sarmento2020musical}. 

Beyond the interest of portraying meaning to users, sonification systems that explore musicality can potentially be used as a creativity tool in the form of an Interactive Musical System (IMS). According to \cite{Wu:2017}, IMSs can offer easy access to the rewarding experience of creating and manipulating sound and music for non-musicians through intuitive and responsive user interfaces. With this same perspective, a few works have combined the use of sonified data with the intent of allowing users to create or manipulate music \cite{Allison2012, Gaye2003, park_sihwa_2010_1177877}.

Contrary to most of the aforementioned papers, the current work combines the three previous concepts by proposing, designing and implementing an IMS capable of generating original music based on urban data. This design idea, however, still applies to a few works \cite{Allison2012, Gaye2003, park_sihwa_2010_1177877}. Out of those, Allison et al. \cite{Allison2012} and Gaye et al. \cite{Gaye2003} aimed to let the user have an \textit{in-situ} exploration of the city, while gathering information close to them to manipulate songs. Our approach, similar to Park et al. \cite{park_sihwa_2010_1177877}, detaches the user from the local environment, enabling them to explore any location on earth via a map. Fundamentally, regarding its data sonification, Park et al. \cite{park_sihwa_2010_1177877} focuses on the data analysis of paths created by users on the map, while our approach sonifies entire areas at once, based on a radial reading of it, similar to a radar. Additionally, Xenakis is meant for producing musical loops, which are intended to be repeated and serve as a basis for music creation.

\section{The Name Xenakis}

In all honesty, this name was decided shortly before the submission, for reasons of exposition. \MYhref{https://miro.com/app/board/o9J_l47ltks=/?moveToWidget=3074457361915705805&cot=14}{Xenakis}' approach to music was very influential in our research and development, and it was clear that he should be discussed and contextualized. Naming the tool after him seemed not only adequate but also quite convenient so we had an excuse to pull him wherever we wanted in the text. Another idea was to reference him indirectly by using a Greek word, since it was his signature in naming compositions: \textit{Metastaseis}, \textit{Pithoprakta}, \textit{Achorripsis}, and so on. We acknowledged that this reference was too obscure for anyone to care.

\section{Conclusion}

This is not over, but there are way too many open possibilities for future work that we need to carefully evaluate how to develop our next steps.  We hope to gain some exposition and foment discussion with this piece, but the process of developing it and being immersed in musical problems has been enriching by itself. Our main contribution, besides the tool itself, is the contextualization of sonification within our visualization background, and \MYhref{https://miro.com/app/board/o9J_l47ltks=/?moveToWidget=3074457361962759559&cot=14}{the development of information theoretical themes from it}. We intend to perform user studies and compare task performance among users with different degrees of musical literacy, and with the visually and hearing impaired. Finally, our miro board, when zoomed out so that all content can be seen, looks like an \MYhref{https://miro.com/app/board/o9J_l47ltks=/?moveToWidget=3074457361977736871&cot=14}{urban map}. It would have been a pretty interesting and meta move to musify the street orientations within it, but the effort was not worth it.

\acknowledgments{
The authors wish to thank musicians Santiago Campos, Werner Silveira, and Elton Oliveira de Souza for shaping this work in different forms. This work was supported in part by
a grant from XYZ.}

\bibliographystyle{abbrv-doi}

\bibliography{template}
\end{document}